\documentclass[12pt,epsf]{article}
\usepackage{epsf}
\usepackage{graphics}

\begin{document}
\title
{May gravitons be super-strong interacting particles?}
\author
{Michael A. Ivanov \\
Physics Dept., \\
Belarus State University of Informatics and Radioelectronics, \\
6 P. Brovka Street,  BY 220027, Minsk, Belarus.\\
E-mail: ivanovma@gw.bsuir.unibel.by.}
%\date{May 8, 2004}
\maketitle

\begin{abstract}
A scheme, in which gravitons are super-strong interacting, is
described. The graviton background with the Planckian spectrum and
a small effective temperature is considered as a reservoir of
gravitons. A cross-section of interaction of a graviton with any
particle is assumed to be a bilinear function of its energies. Any
pair of bodies are attracting not due to an exchange with its own
gravitons, but due to a pressure of external gravitons of this
background. A graviton pairing is necessary to obtain classical
gravity. Any divergencies are not possible in such the model
because of natural smooth cut-offs of the graviton spectrum from
both sides. Some cosmological consequences of this scheme are
discussed, too. Also it is shown here that the main conjecture of
this approach may be verified at present on the Earth.
\end{abstract}

\section[1]{Introduction }
What is a quantum mechanism of gravity? To answer this question,
it is necessary to keep in mind a few different circumstances. A
commonly accepted hypothesis, that an exchange with gravitons,
which are radiated by bodies itself, causes classical gravity, is
not a single possible one - gravitons might belong to an external
sea of particles which exists independently of attracting bodies.
A coupling constant of quantum gravitational interaction would
differ from the Newton constant and may depend on energies of
interacting particles. A geometrical language for short distances
may not be adequate to describe quantum gravity. Perhaps, on the
present stage when we know so a little about this mechanism, it
would be better to consider a problem of its searching as a
separate one. Some known effects which are not usually connected
with gravity may be involved in a circle of interests of
researchers by considerations of possible mechanisms. At first, it
concerns cosmological effects which manifest themselves only on
huge distances and during big times.
\par
There are some facts beyond cosmology, which should be taken into
account in this search, too. The Pioneer 10 anomaly \cite{4} is
one of them. Another fact is an observation of discrete energy
states of neutrons in the Earth's gravitational field by
Nesvizhevsky et al. \cite{11}. In this remarkable experiment we
see the huge difference - about 40 orders - between observed state
energies $\sim 10^{-12}$ eV and the Planck energy of $\sim
10^{19}$ GeV which is expected from dimension reasonings as a
threshold of any quantum gravity effect. The long known
contradiction between the general relativity and quantum mechanics
in descriptions of a microparticle motion is the third fact: if in
one theory all particles should move on geodesics, in another they
cannot move on definite trajectories. Maybe, a cause of this
contradiction is that both theories do not take into account
influences of single gravitons on a microparticle; then small
graviton energies in a future theory are appeared to be more
appropriate than the Planchian ones. A possible compositeness
\cite{15} of the fundamental fermions - electrons, neutrinos,
quarks etc - should be taken into account, too. Components of
these composite fermions would be bounded with a quantum
gravitational interaction which is not similar to ordinary
gravity. \par The main features of a quantum model of classical
gravity and its cosmological consequences are described here. The
model is based on the conjecture about an existence of the
background of super-strong interacting gravitons.

\section[2]{A gravitational attraction due to the background}
The important features of this model are the following ones (for
more details, see \cite{4b,4c}).

$\\\bullet$ The graviton background has the Planckian spectrum and
the same temperature $T$ as CMB. $\\\bullet$ The graviton
background is in a state of dynamical equilibrium: it is cooled
via self-interactions of gravitons and formation of virtual
massive gravitons which may be dark matter particles, and is
heated up via interactions with other radiations \cite{16}.
$\\\bullet$ A cross-section of interaction $\sigma (E,\epsilon)$
of a graviton with any particle is a bilinear function of its
energies: $\sigma (E,\epsilon)= D \cdot E \cdot \epsilon,$ where
$D$ is some new dimensional constant, $E$ is a particle's energy,
$\epsilon$ is a graviton's energy. The estimate of $D$ is: $D \sim
10^{-27} m^{2}/eV^{2},$ i.e. gravitons are super-strong
interacting particles. $\\\bullet$ Due to a pressure of single
gravitons, there act equal attractive and repulsive forces of
three order greater than the Newtonian force between any two
bodies, but a net force is equal to zero. $\\\bullet$ To ensure
Newtonian attraction, a pairing of single gravitons of running
flux is necessary, and such pairs should be destructed by
collisions with a body. A nature of this pairing remains unknown.
If this pairs have spin 2, then single gravitons may have spin 1.
Only two modes of spin-2 particles may exist in the model.
$\\\bullet$ Given this pairing, the Newton constant $G$ is equal
to: $$ G \equiv {2 \over 3} \cdot {D^{2} c(kT)^{6} \over
{\pi^{3}\hbar^{3}}} \cdot I_{2},$$ where $k$ is the Boltzmann
constant, $I_{2} = 2.3184 \cdot 10^{-6}.$ $\\\bullet$ In the case
of interaction of gravitons with big bodies in the aggregate, it
is impossible to have Newton's law. One needs an "atomic
structure" of matter to get this law. $\\\bullet$ For proton-mass
particles, the equivalence principle should be broken at distances
$\sim 10^{-11} \ m,$ if the model is true. It means that at
shorter distances gravity cannot be described alone, without other
known interactions. It is also the limit to apply a geometrical
language in gravity.

\section[3]{Cosmological consequences of the model}
Such the mechanism of gravity should have the following
cosmological consequences \cite{1,4b}: $\\\bullet$ A quantum
interaction of photons with the graviton background would lead to
redshifts of remote objects; the Hubble constant $H$ is equal in
this model to: $H= {1 \over 2\pi} D \cdot \bar \epsilon \cdot
(\sigma T^{4}),$ where $\bar \epsilon$ is an average graviton
energy, $\sigma$ is the Stephan-Boltzmann constant. Redshifts are
caused by forehead collisions with gravitons. $\\\bullet$ The
Hubble constant is connected in this approach with the Newton one
as:
$$H= (G  {45 \over 32 \pi^{5}}  {\sigma T^{4} I_{4}^{2}
\over {c^{3}I_{2}}})^{1/2}= 3.026 \cdot 10^{-18}s^{-1},$$ where
$I_{4}=24.866.$ $\\\bullet$ Due to non-forehead collisions with
gravitons, an additional relaxation of any photonic flux leads to
the luminosity distance $D_{L}:$
$$D_{L}=a^{-1} \ln(1+z)\cdot (1+z)^{(1+b)/2},$$
where $a=H/c,$ $z$ is a redshift, and the relaxation factor $b$ is
equal to \cite{7}: $b= 3/2+2/\pi =2.137.$ See a comparison of this
function with observations of \cite{12} in my paper \cite{14}.
$\\\bullet$ Any light radiation spectrum will be deformed due to
the quantum nature of red-shifting process. A theory of this
effect does not exist today. But it may be checked experimentally
in a laser experiment (see the next section). $\\
\bullet$ Any massive objects, moving relative to the background,
should be decelerated by the background. A body's acceleration $w$
by a non-zero velocity $v$ relative to the background is equal to:
$w = - ac^{2}(1-v^{2}/c^{2}),$ and has by small velocities the
same order $Hc$ as an anomalous acceleration
of Pioneer 10 \cite{4}. $\\
\bullet$ An existence of black holes contradicts to Einstein's
equivalence principle in a frame of this model \cite{4b}.

\section[3]{How to verify the main conjecture of this approach}
I would like to show here a full realizability at present time of
verifying my basic conjecture about the quantum gravitational
nature of redshifts in a ground-based laser experiment. Of course,
many details of this precision experiment will be in full
authority of experimentalists. \par It was not clear in 1995 how
big is a temperature of the graviton background, and my proposal
\cite{111} to verify the conjecture about the described local
quantum character of redshifts turned out to be very rigid: a
laser with instability of $\sim 10^{-17}$ hasn't appeared after 9
years. But if $T=2.7 K$, the satellite of main laser line of
frequency $\nu$ after passing the delay line will be red-shifted
at $\sim 10^{-3}$ eV/h and its position will be fixed, but, due to
the quantum nature of shifting process, the ratio of its intensity
to main line's intensity should have the order: $$\sim {h\nu \over
\bar{\epsilon}}{H\over c} l,$$ where $l$ is a path of laser
photons in a vacuum tube of delay line. It gives us a possibility
to plan a laser-based experiment to verify the basic conjecture of
this approach with much softer demands to the equipment. An
instability of a laser of a power $P$ must be only $\ll 10^{-3}$
if a photon energy is of $\sim 1~eV$. If one compares intensities
of the red-shifted satellite at the very beginning of the path $l$
and after it, and uses a single photon counter to measure the ones
(when $q$ is a quantum output of a cathode of the used
photomultiplier, $N_{n}$ is a frequency of its noise pulses, and
$n$ is a desired ratio of a signal to noise's standard deviation),
then an evaluated time duration $t$ of collecting data would have
the order: $$ t= {\bar{\epsilon}^{2}c^{2} \over H^{2}} {n^{2}N_{n}
\over q^{2} P^{2} l^{2} }.$$ Assuming $n=10,~N_{n}=10^{3}
~s^{-1},~ q=0.3, ~P=300~ W,~ l=100 ~m, $ we get: $t \sim 4$ days,
that is acceptable for the experiment of such the potential
importance.

\section[4]{Conclusion}
The described model of Le Sage's kind has not an analogue in
present-day physics of particles. If this mechanism is realized in
the nature, both the general relativity and quantum mechanics
should be modified. The indirectly observed objects in centers of
galaxies, which are known now as black holes, should have another
nature, too. Gravity at short distances, which are meantime much
bigger than the Planck length, needs to be described only in some
unified manner.

\end{document}